\documentclass[conference]{IEEEtran}
\IEEEoverridecommandlockouts
\usepackage{cite}
\usepackage{amsmath,amssymb,amsfonts}
\usepackage{algorithmic}
\usepackage{graphicx}
\usepackage{subcaption}
\usepackage{textcomp}
\usepackage{xcolor}
\usepackage{hyperref}
\def\BibTeX{{\rm B\kern-.05em{\sc i\kern-.025em b}\kern-.08em
    T\kern-.1667em\lower.7ex\hbox{E}\kern-.125emX}}
\begin{document}

\title{\vspace{18pt}Towards Reproducible Evaluations for Flying Drone Controllers in Virtual Environments \\
%
}

\author{\IEEEauthorblockN{Zheng Li}
\IEEEauthorblockA{\textit{Fudan University} \\
Shanghai, China} 
\and
\IEEEauthorblockN{Yiming Huang, Yui-Pan Yau, Pan Hui}
\IEEEauthorblockA{
\textit{The Hong Kong University of Science and Technology}\\
HKSAR, China} 
\and
\IEEEauthorblockN{Lik-Hang Lee*}\thanks{*Lik-Hang Lee is the corresponding author. Z. Li and Y. Huang were interns during the study of this work with KAIST. The primary affiliation of Pan Hui is The Hong Kong University of Science and Technology (Guangzhou), China. He is also affiliate with the University of Helsinki, Finland. Acknowledgment: This work has been supported by the Academy of Finland, under Grant 319669, and Grant 325570.}
\IEEEauthorblockA{
\textit{KAIST}\\
Daejeon, South Korea} 
}

\maketitle

\begin{abstract}
Research attention on natural user interfaces (NUIs) for drone flights are rising.
Nevertheless, NUIs are highly diversified, and primarily evaluated by different physical environments leading to hard-to-compare performance between such solutions. 
We propose a virtual environment, namely VRFlightSim, enabling comparative evaluations with enriched drone flight details to address this issue. We first replicated a state-of-the-art (SOTA) interface and designed two tasks (crossing and pointing) in our virtual environment. Then, two user studies with 13 participants demonstrate the necessity of VRFlightSim and further highlight the potential of open-data interface designs. 
\end{abstract}

\begin{IEEEkeywords}
Human-Drone Interaction, Methodologies, Virtual Environments, The Fitts's Law, Metaverse.
\end{IEEEkeywords}

\section{Introduction}

In recent years, quadcopter-style drones have been widely available due to their highly portable features, and users can control their drones with dedicated controllers. Drones can serve as a representative in our physical world, and users with augmented reality headsets, by leveraging the drone views, can teleport to such environments~\cite{pinpointfly} for various industrial and commercial purposes~\cite{AR-IROS}\cite{d9-DroneCHI19}. 
Although the two-handed remote control transmitters (RCT) dominate the way of drone flights, one-handed controllers, e.g., commercial products: \textit{FT Aviator} and research prototypes: \textit{DroneCTRL}~\cite{DroneCTRL}, allows users to enjoy higher mobility by reserving a spare hand for other tasks, e.g., holding a handrail in a train. 
Furthermore, researchers have explored alternative solutions to pursue more natural human-drone interaction~\cite{bodyGesture-drone} that promotes invisible interfaces inherited in our bodies, e.g., gaze-driven drone flights~\cite{d7-gaze} and ring-form addendum~\cite{arthur-drone}. However, the proliferation of flying drone controllers leads to difficult comparisons between research prototypes. Additionally, the performance of such controller prototypes is subject to flying tasks in real-world environments that include a collection of external factors, such as physical obstacles, and weather conditions (e.g., windy or rainy days), More notably, replicating research prototypes and flying tasks are usually time-consuming and expensive. 

\begin{figure}[htp]
     \centering
     \begin{subfigure}[b]{0.45\linewidth}
         \centering
         \includegraphics[width=\textwidth]{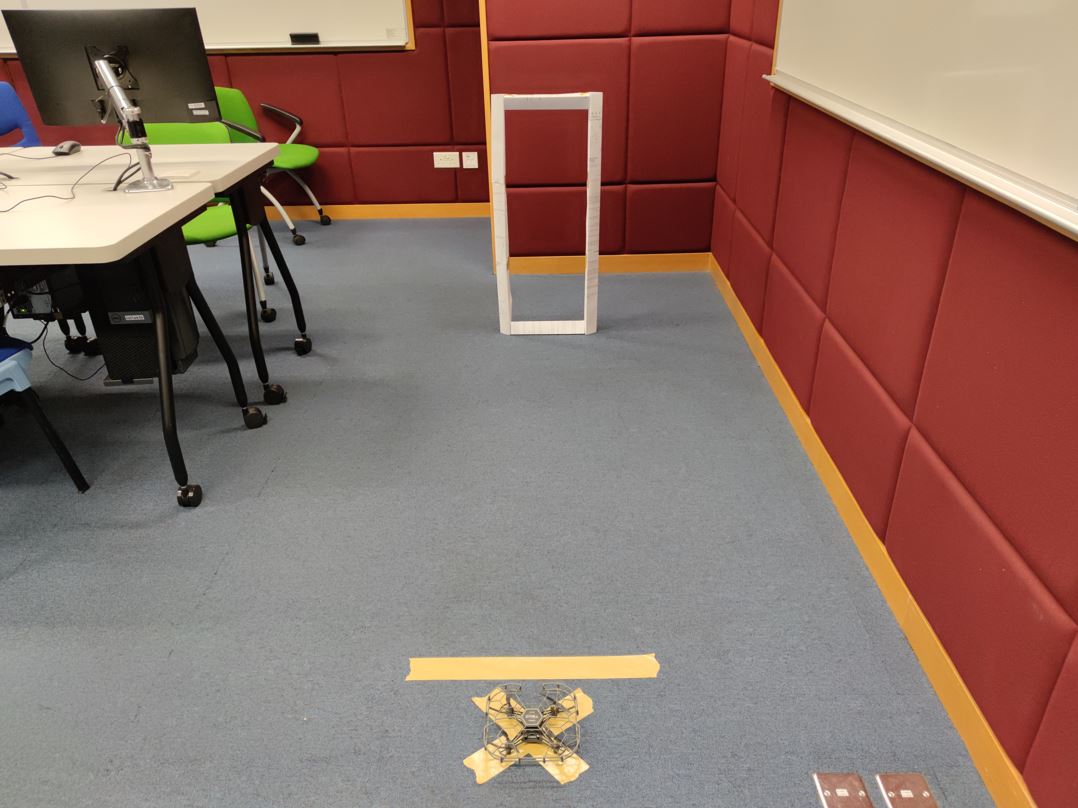}
         \caption{Crossing task}
         \label{fig:crossing-velocity}
     \end{subfigure}
     \begin{subfigure}[b]{0.45\linewidth}
         \centering
         \includegraphics[width=\textwidth]{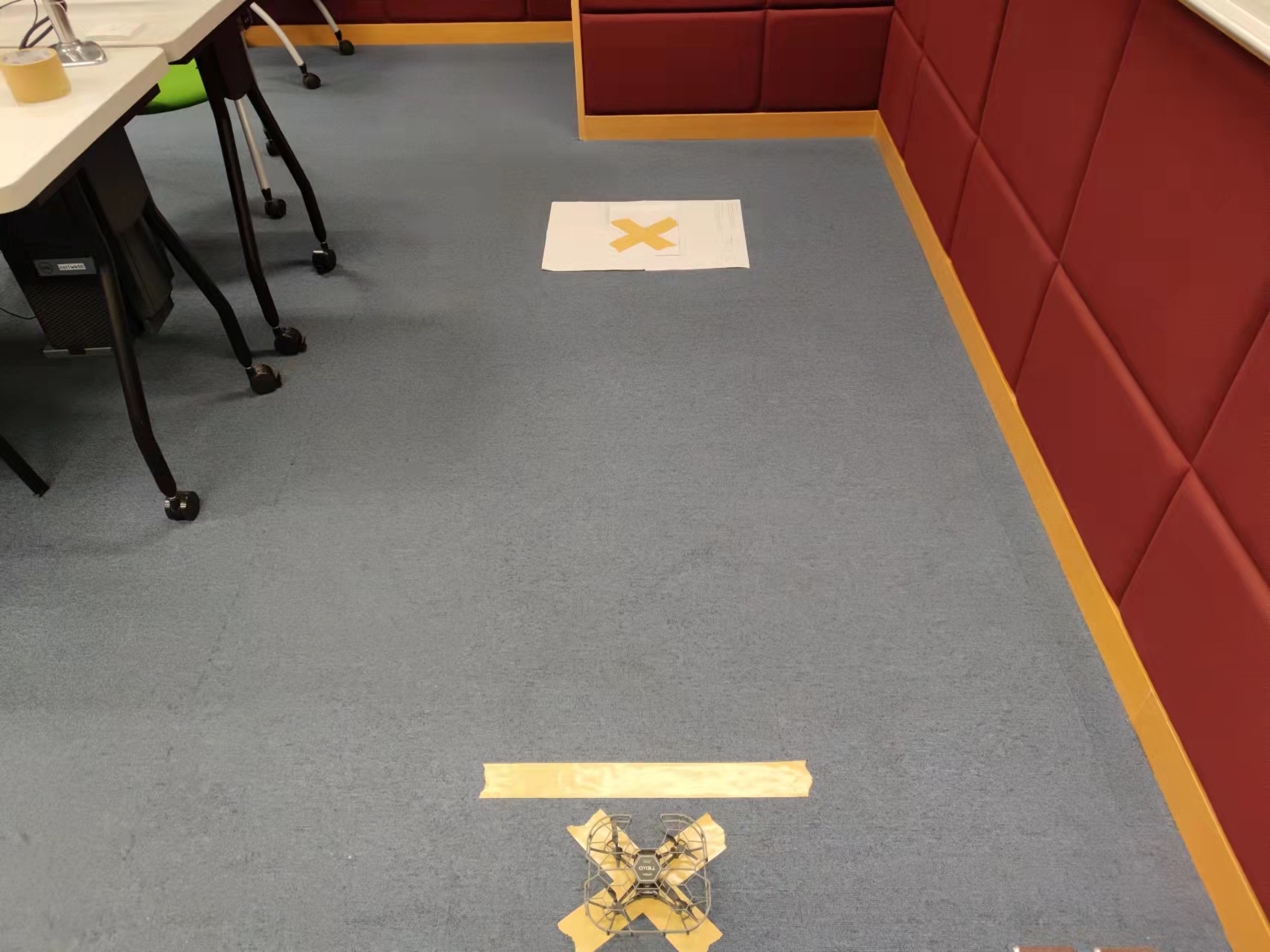}
         \caption{Pointing task}
         \label{fig:crossing-acceleration}
     \end{subfigure}
        \caption{Drone interaction experimental tasks in our physical lab that are space-occupying and time-consuming.}
    \centering
    \label{fig:teaser}
\end{figure}

Therefore, we leverage virtual environments (Microsoft Airsim) to develop an evaluation tool that offers comprehensive profiles of controller prototypes. 
Our virtual environment enables interaction designers to create new tasks (e.g., flying routes, checkpoints, and obstacles), and hence records both the performance of users and flying drones. In our current tool, the performance refers to users' task completion time, as well as drones' velocity, acceleration, jerk, and flying trajectory. 
A newly designed controller can be evaluated by a task in which a flying drone reaches checkpoints in a high-resolution virtual world. 
It is worth noting that constraints of task design appear with traditional measurements in physical worlds (Figure~\ref{fig:teaser}), including short and simple task~\cite{arthur-drone}, manual measurements~\cite{Yamada2019ModelingDC}, room-size setup, short battery life and hence long recharging time~\cite{pinpointfly}. In contrast, conducting evaluations in virtual environments can potentially get rid of the aforementioned constraints, hence facilitating the evaluation progress. 

To illustrate the prominent features of our evaluation tool, we implemented a prototypical interface of a SOTA controller~\cite{arthur-drone}, driven by one-handed and one-thumb operations, and further compared the SOTA interface with a baseline that emulates a two-handed tangible controller. Our evaluation tool reveals new insights into both the two-handed and one-handed controllers, in which two-handed controller remains superior to the SOTA method in complicated scenarios. Interaction designers can utilize this design cue to rationalize the practical use cases in physical environments. To conclude, this paper primarily contributes to an evaluation tool for drone controllers with comprehensive metrics that offers comparable results between controller prototypes, facilitating result replicability and open-data research. 

\section{Related Works}
This section highlights the related work of aircraft control, simulation, and virtual environments.
As one of the most widely accessible robotic crafts to the public, drones have attracted many researchers' attention. Different controllers are developed to enhance human-drone-interaction scenarios, including the traditional hand movement controller~\cite{obaid2016would}, eye-gaze-based controller~\cite{hansen2014use}, voice-based-controller~\cite{landau2017system}, etc. However, all of these controllers should be tested in a standardized environment, which takes great effort to deploy. 

Experiments for robotic deployment in physical environments calls for resources of time and space. Thus, simulation platforms have been built, allowing easy-to-access, faster and low-cost evaluations, for instance, behavioral dynamics between human avatars and robotics~\cite{10.1145/2559636.2559841}, simulation platforms enable emulations of 3D-printed robots~\cite{9636114}, an interactive robotic arm for housework~\cite{9636667}, social robot navigation~\cite{9636319}, etc. Lately, evaluation of commercially-available service robots moved to online platforms for scalable evaluations. 
On the other hand, the emerging AR/VR in recent years can serve as an alternative to simulation. The virtual environment can go beyond physical constraints. For instance, AR can help architects to understand the user perception of movable walls~\cite{HRI-building-robot}, while building such movable walls in real buildings are usually impractical. Similarly, mixed reality robots can induce people to empathize with bad incidents~\cite{10.1145/2696454.2696471}. VR environments can deliver a testing ground of risky operations in which injury and casualty are not affordable, for instance, virtual driving and risky driving events~\cite{10.1145/3173574.3173739}, and virtual jumping between intervals to observe users' locomotion~\cite{10.1145/3313831.3376243}.

We acknowledge that numerous open-source drone flight simulators exist, e.g., autonomous drones~\cite{swarm} and training modules for human operators~\cite{training-sim}. However, a recent survey~\cite{Ebeid2018ASO} highlights the difficulty of evaluating one flight controller, i.e., substantially high development time and testing efforts. 
Also, our work uniquely leverages virtual environments to resolve the lack of standardization of flight controller architectures and hence, drone controllers' evaluation.

\section{System Architecture}


Our system aims to provide a ubiquitous and open-data environment for human-drone interaction research, with a focus on the designs of drone controllers. We implement the system with four significant modules in order to create a standardized environment for conducting the human-drone experiment systematically (Figure \ref{fig:system-arch}). As mentioned, one of the limitations of drone experiments is the difficulty of efficiently comparing results among the research community. To resolve this problem, we aggregate the data from Input, Pathway, and Environment modules into experiment logs for a reproducible and trackable experiment. Another challenge is the external factors of drone experiments, for instance, the drone models, drone sizes and speeds, and the physical environment: wind speed and weather. For manipulating the external contributors, we implement the Environment module and Drone state manager indirectly linked to the Airsim platform~\cite{shah2017airsim}, based on the UNREAL engine. Therefore, we can standardize all the experimental factors and perform human-drone interaction experiments in an identified yet united environment.

\subsection{System Components}

Utilizing the Airsim binary (editable UNREAL project), we implement the experimental evaluation system (Figure~\ref{fig:system-arch}) with extra programmed modules for manual drone control, network communication protocols, and input interfaces. 1) \textit{Manually drone control.} The extra functionality for controller interface support. 2) \textit{network communication protocols. Information exchanging service between input devices and evaluation platform. 3) \textit{Input interfaces.} GUIs for adjusting experiment settings and filling in personal information. We create a portable released package based on the original Airsim block binary by removing most redundant assets and adding the necessary experiment objects. With a combination of code and binary,} we construct an experimental platform without re-compiling the Airsim source code and re-installing the UNREAL engine. Besides the code and binary, the system also includes the 3D asset files for two primary operations: pointing and crossing~\cite{Yamada2019ModelingDC}\cite{modelling-fitts}. 
The checker asset is a plate for the pointing operation, and the crossing asset is a door frame for the crossing operation. A drone path exists for each pointing and crossing operation, including start points, endpoints, and checkpoints. The nine major programmed modules are as follows: 
\begin{itemize}
    \item \textbf{Controller-Keyboard}: Directing drone's movements with the controllers (e.g., keyboards/touchscreens). 
    \item \textbf{Drone Environment Parameter Controller}: Storing the environment setting.
    \item \textbf{Drone Environment Setting GUI}: User interface for adjusting the environment setting. 
    \item \textbf{Drone State}: Storing the drone state setting. 
    \item \textbf{Experiment Logger}: Logging experiment results.
    \item \textbf{Main Program}: The executable file for experiments that initializes the whole experiment. 
    \item \textbf{Path Generator}: Generating the experimental task -- drone flying path. 
    \item \textbf{Path Manager}: storing the setting of the experiment path.
    \item \textbf{UDP Server}: UDP communication between devices.  
\end{itemize}

\subsection{Component interaction}

\begin{center}
\begin{figure}
  \includegraphics[width=\linewidth]{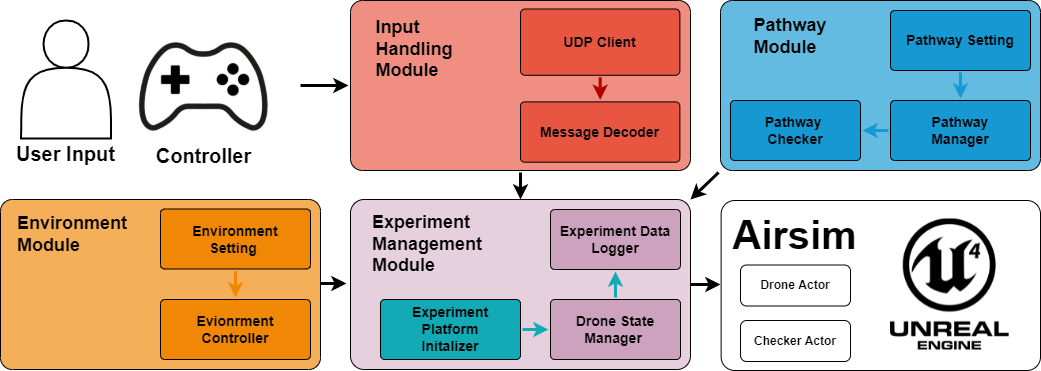}
  \caption{System architecture}
  \label{fig:system-arch}
\end{figure}
\end{center}

These modules of the experiment platform are independent and only exchange the state and setting information (Figure \ref{fig:system-arch}) between each other. The executed Main Program is the container for the components, including Drone State, Path Manager, UDP Server, Controller, Environment Parameter Controller, and Drone Environment Setting GUI. The drone state module passes the Airsim platform client through the MAIN to other components. The drone state module also provides the drone velocity, angle, acceleration, position, and events that record the drone collision. The experiment data flows from the Drone State component to the Experiment Logger, Path Manager, and Drone Environment Parameter Controller. 

The Environment Setting GUI passes the participant's message and the input configuration to the Environment Parameter Controller and simultaneously the ``Start" and ``End" commands to the UDP Server components. The drone state message and the experiment event will occur when the Drone State takes the corresponding command(s) for poi from the UDP server. For example, hitting the door frame and plate checker notifies the Path Manager, then the Path Generator changes the state of the checker. Meanwhile, the Drone State component will call the APIs of the Experiment Logger, thus logging the event(s) and the Drone State message(s). 

\section{Experiment Design}
To understand the efficacy of VRFlightSim for evaluating controllers with 
Human-Drone interaction, we implemented two common drone operations, namely 
drone pointing and crossing operations on the Microsoft Airsim platform (version: 1.5.0.) \cite{shah2017airsim} with UNREAL Engine (version: 4.25.4) 
The pointing operation refers to a drone flying from one location to another, and both start and end points require the drone to take off and land, respectively~\cite{modelling-fitts}. The crossing operation refers to a drone flying through a door frame in mid-air~\cite{Yamada2019ModelingDC}, which can be regarded as a subset of the pointing operation. 

\subsection{Participants and Apparatus}
We recruited 
13 participants (19 – 28 years old; 11 male and 2 female) to conduct the drone interaction experiment with VRFligthSim. All participants, who are experienced smartphone users, are students from our university campus. Only two of them have experience with drone flights, while the others are inexperienced, i.e., first-time drone users. The participation is based on informed consent and is truly voluntary. The experimental procedures comply with the regulations of GDPR and the IRB of our universities. All user-generated data were de-identified and password-protected, and will be deleted after the project completion.
\begin{figure}[h]
    \centering
    \includegraphics{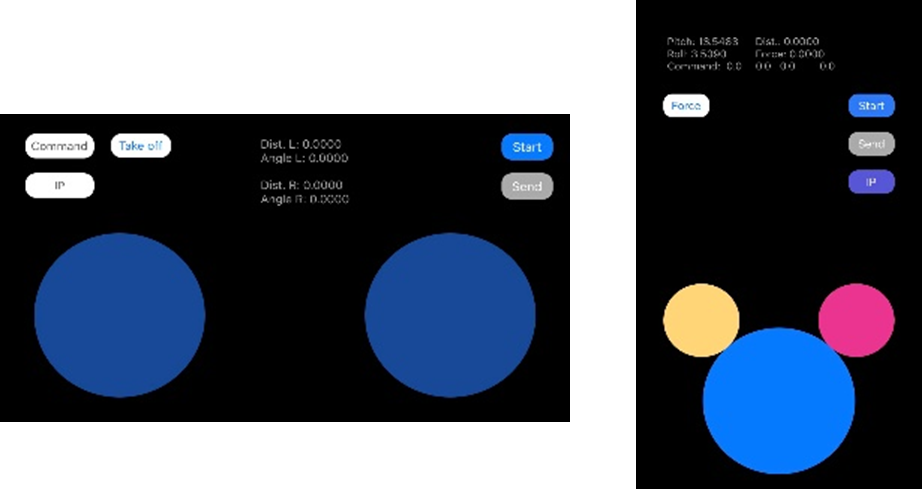}
    \caption{Controller interfaces of Two-button (Baseline, left) and One-handed (SOTA, right) operations on the blue button(s). }
    \label{fig:freakingbusy}
\end{figure}

We employed an iPhone 7 that allows us to replicate the SOTA control interface, driven by the 3-D touch function (i.e., force detection), that resulted in a ring-form controller~\cite{arthur-drone}, and a baseline that emulates a tangible two-button controller (see Figure~\ref{fig:freakingbusy}). \textcolor{black}{We chose these touchscreen-based controllers as typical methods, as smartphones support many drone commodities in recent years.}
For the SOTA control interface, the drone speed is determined by the force exertion level of the user, given by the force sensor on iPhone 7~\cite{arthur-drone}. Drone speeds for the baseline interface is derived from the distance between the touch point and the mid-point of the button.
The smartphone connects to a computer running the VRFlightSim via WIFI, which sends the RC command for controlling the simulated drone by the UDP client programmed in Swift (IOS). Also, the received RC commands are handled by a UDP server programmed in Python on the computer. The received RC commands and the corresponding actions of the drone will first go through the API of VRFlightSIM (written in Python), and then be processed by the API of Airsim, based on the UNREAL engine. 

\subsection{Experimental Setup in the Virtual Environment}
\begin{figure}[htp]
     \centering
     \begin{subfigure}[b]{0.45\linewidth}
         \centering
         \includegraphics[width=\textwidth]{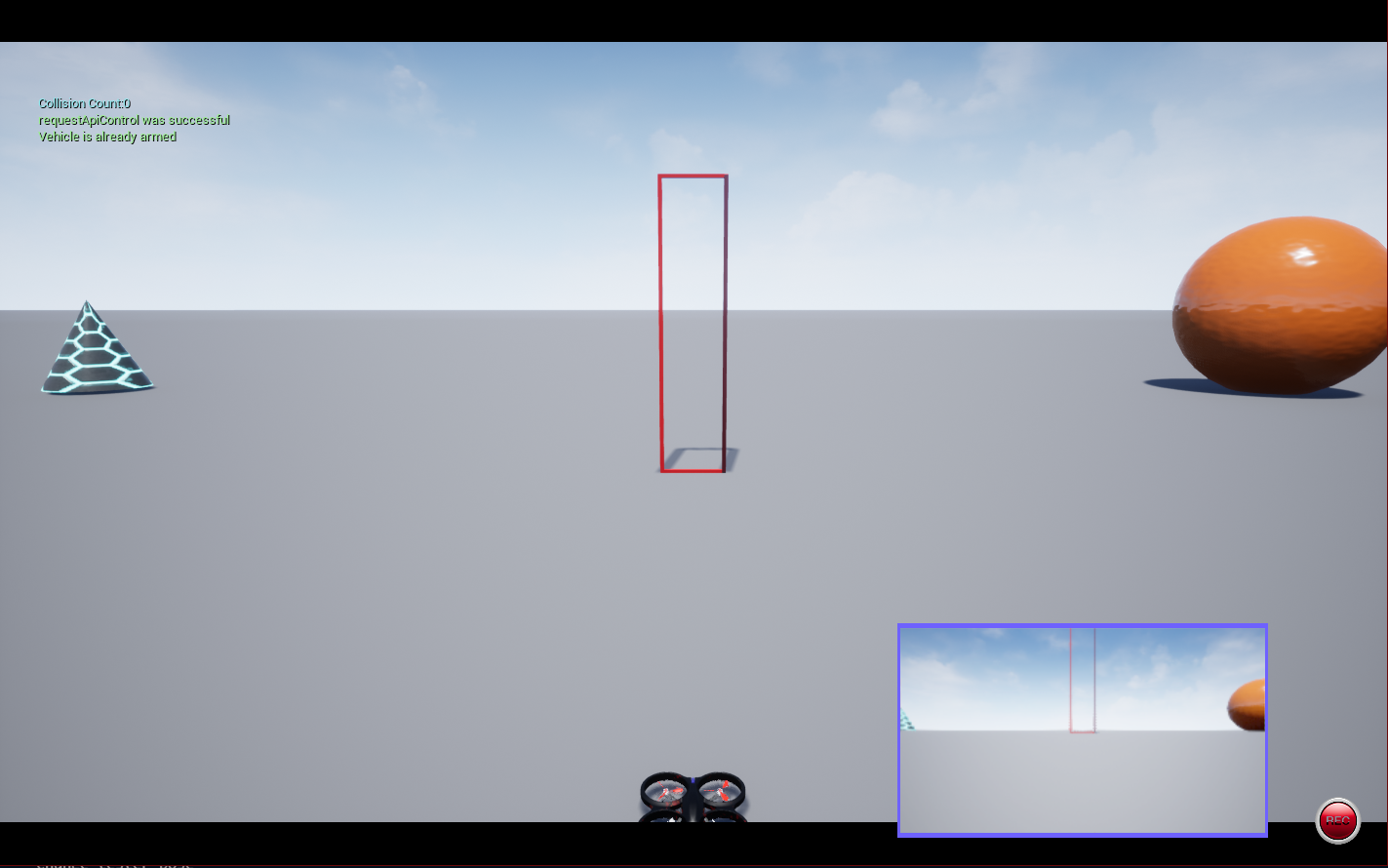}
         \caption{Crossing task}
         \label{fig:cross-vr}
     \end{subfigure}
     \begin{subfigure}[b]{0.45\linewidth}
         \centering
         \includegraphics[width=\textwidth]{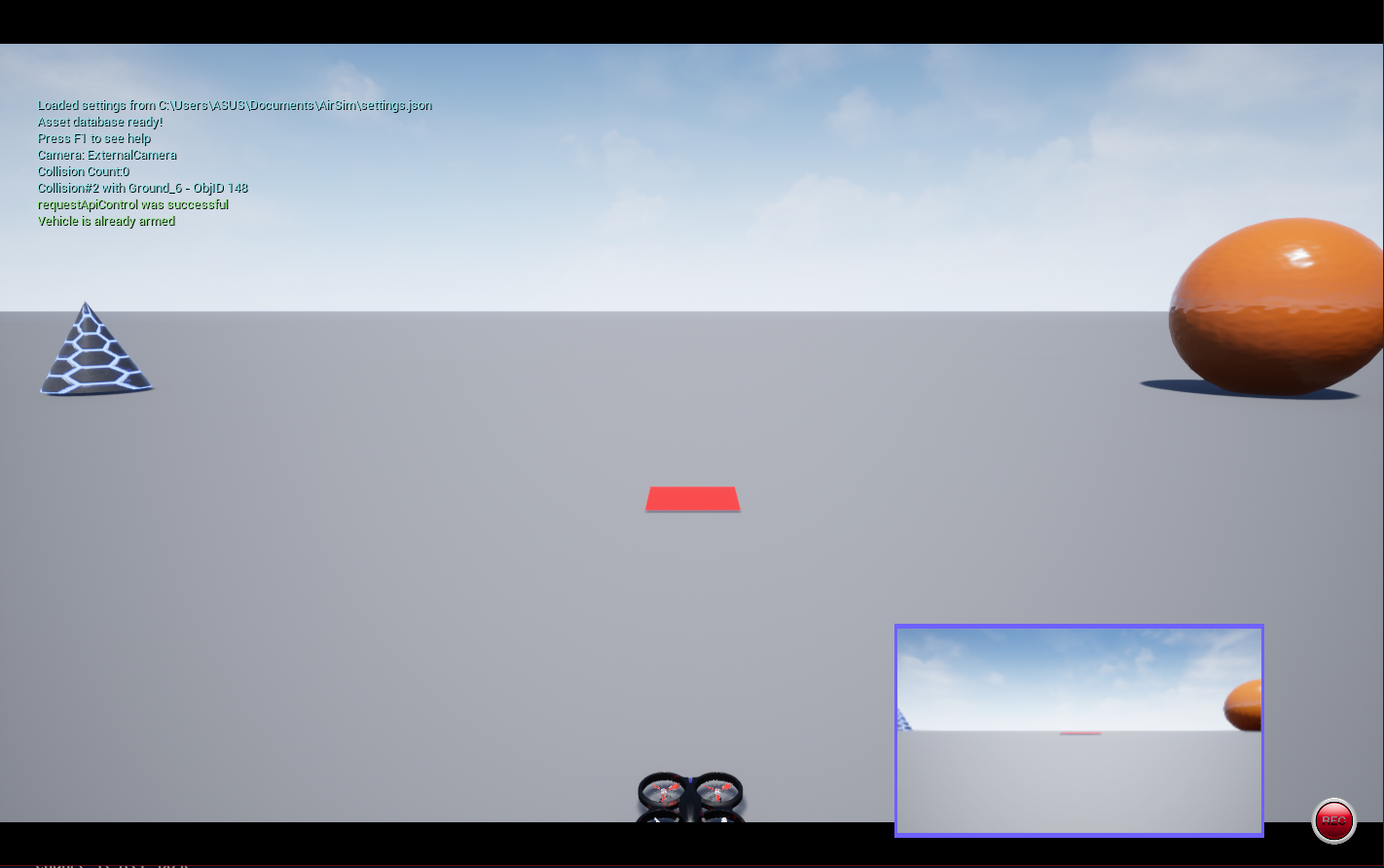}
         \caption{Pointing task}
         \label{fig:pt-vr}
     \end{subfigure}
        \caption{Experiments Tasks for Drone Flights in VRFlightSim.}
    \centering
    \label{fig:setup}
\end{figure}

We re-implement the settings of pointing and crossing tasks, based on prior works~\cite{modelling-fitts}\cite{Yamada2019ModelingDC} in our virtual worlds.
The user perspective in the Airsim environment is set as a 90° field-of-view and 165cm height that emulates the normal vision of a person with the average height in our population. As such, the position of the camera is fixed at the starting point (0, 0, 1.65). The weather of the experimental environment defaults as ``sunshine" initially, which corresponds to a satisfactory level of illumination, and the wind level of the experiment block is zero (i.e., turn off). 
We did our utmost to minimize the distraction from the background and the checkpoints.
Hence, as shown in Figure~\ref{fig:setup}, we employ a plain colour background, in addition to two objects (a triangular pyramid and a cube) to assist the participants in recognizing the directions. Also, the experimental checkers (i.e., checkpoints) are built using the basic box asset in the UNREAL engine with the modification in size and colour. The pointing checker is a red-colour square plate, and the crossing checker is a red-colour and squared door frame perpendicular to the ground with blank space in the door frame facing the participants. The middle part of the crossing checker is a piece of transparent box-sized material that serves as the trigger for the mission’s completion. Finally, the drone model in this experiment is a quadcopter named “Parrot Mambo Fly”, sized at 0.18 x 0.18 (m), that replicates the experimental setting for casual drone flights~\cite{arthur-drone}. 

\subsection{Task and Procedures}

For each operation type, either crossing or pointing, the participants ran five trials with TWO types of controllers (two-button baseline and one-handed, under the virtual tasks having THREE different sizes of experimental checker (Landing areas for pointing: .4m x .4m, .7m x .7m, and 1.1m x 1.1m; Mid-air door frames for the crossing: .3m x .3m, .4m x .4m, and .5m x .5m) and TWO flying distances (Pointing: 2m and 4m; Crossing: 2.5m and 3,5m). As a result, the experiment setting is regarded as a 2 controllers * 3 difficulties * 2 distances, for two operations with five repetitions. 
The five-trial runs are sufficient, as the extension of the trial number will lead to negative performance on the participants, while the tiredness and boredom of the participant lead to users' burnt out.
The randomized orders of tasks and participants follow~\cite{d53-10.5555/2501707} to reduce the  asymmetric skill transfers, and hence the threat of internal invalidity, i.e., carry-over effects. 

At the beginning of the test, we asked every participant to watch a video demonstrating the flight task performed by an experienced pilot for both the pointing and crossing tasks. Afterward, they made some trials, no longer than five minutes to get basic ideas about the interfaces and the virtual environments. The five-minute session is sufficient for the participants to get familiar with the basic control of the drone and understand the task requirement of the experiment. Next, the participants ran the formal experimental trials. They had been told to finish the task as fast and as accurately as possible. 

We require the participants to press the start button on the iPhone interface, and subsequently drive the drone to the target with the minimum time they can achieve. After reaching the target, the participants will get a completion message from the PC display, and the participants require to press the stop button to end the current trial. 
After ending a trial, a file saving message for the RC commands recording on the iPhone is shown, and the researchers saved this file to a directory with their name and the testing conditions. 
If a trial flight fails due to the mistaken operations caused by the participants or the program crash, we simply neglect the experiment result(s) and reran another trial.  
We collected all IMU data, including the timestamp with the coordinate(s), velocity and acceleration values of all directions, and collision events with the coordinate of the contact point between the drone and collided objects. The Airsim API supports all the aforementioned data collection pipelines at a granularity of every 10ms. 
The collected data were analyzed for generating the trajectories of drone flights, the completion times of all experimental conditions, as well as the illustration of the predictive models of linear regression based on the Fitts’s Law~\cite{Yamada2019ModelingDC}\cite{modelling-fitts}.

\section{Evaluation Results (Case Study)}

This section first highlights the results of crossing operations and subsequently pointing operations with all the metrics. All the participants' performance with two kinds of controllers under different operations are parsed by three high-level perspectives: completion time, drone states (Velocity, Acceleration, and Jerk), as well as trajectory areas. We depict all the statistical analyses (two-way ANOVA), in Tables~\ref{table:crossing-anova} and~\ref{table:pointing-anova}, for the crossing and pointing operations, respectively.

All the graph plots in this section follow the difficulty index (i.e., ID on the x-axis), inspired by Fitts's Law~\cite{Yamada2019ModelingDC}. 
The original definition of Fitts's law states that the amount of completion time it takes for the device moving from the current position to the target is highly related to the width of the target\((W) \) and the distance between two points\((D) \). To be more precise, the completion time is proportional to \(log(2D/W) \) (i.e., Fitts's index of difficulty), which has been commonly used to determine whether a controller is easy to handle in the scenario of human-computer interaction. 

We collect all the experiment results of crossing \& pointing tasks, and draw the linear relationship of dependent variables (e.g., 
time) in each trial and review the performance in different difficulties. 
Note that the larger the proportional coefficient of the ID is, the more challenging users could handle the drone flight 
as the tasks' difficulty grows. 

\begin{table*}[h]
\centering
\begin{tabular}{|c|cc|cc|cc|cc|cc|}
\hline
\textbf{Metric}    & \multicolumn{2}{c|}{\textbf{Completion Time}}                 & \multicolumn{2}{c|}{\textbf{Velocity}}                        & \multicolumn{2}{c|}{\textbf{Acceleration}}                    & \multicolumn{2}{c|}{\textbf{Jerk}}                            & \multicolumn{2}{c|}{\textbf{Trajectory Area}}                 \\ \hline
\textbf{Factor}    & \multicolumn{1}{c|}{\textbf{p-value}} & \textbf{F-statistics} & \multicolumn{1}{c|}{\textbf{p-value}} & \textbf{F-statistics} & \multicolumn{1}{c|}{\textbf{p-value}} & \textbf{F-statistics} & \multicolumn{1}{c|}{\textbf{p-value}} & \textbf{F-statistics} & \multicolumn{1}{c|}{\textbf{p-value}} & \textbf{F-statistics} \\ \hline
\textbf{Mode}      & \multicolumn{1}{c|}{5.80E-08}         & 30.0726               & \multicolumn{1}{c|}{1.39E-38}         & 205.0140              & \multicolumn{1}{c|}{5.28E-50}         & 285.8583              & \multicolumn{1}{c|}{4.82E-04}         & 12.3645               & \multicolumn{1}{c|}{6.08E-11}         & 44.7612               \\ \hline
\textbf{ID}        & \multicolumn{1}{c|}{1.41E-08}         & 9.2844                & \multicolumn{1}{c|}{3.07E-03}         & 4.0534                & \multicolumn{1}{c|}{0.0931}           & 2.0031                & \multicolumn{1}{c|}{4.65E-08}         & 10.3923               & \multicolumn{1}{c|}{2.59E-07}         & 8.0527                \\ \hline
\textbf{Mode * ID} & \multicolumn{1}{c|}{0.2955}           & 1.2251                & \multicolumn{1}{c|}{0.1808}           & 1.5715                & \multicolumn{1}{c|}{0.5801}           & 0.7177                & \multicolumn{1}{c|}{0.8200}           & 0.3841                & \multicolumn{1}{c|}{0.4284}           & 0.9816                \\ \hline
\end{tabular}
\caption{Two-way ANOVA analysis of Crossing Tasks: DoF = (1,5). All the Controller Interfaces (Mode) and Index of Difficulties (ID) passed the threshold of p-value$<$0.05, except ID for acceleration (0.093).}
\label{table:crossing-anova}
\end{table*}

\subsection{Crossing}
Different from the pointing task, the crossing task is a sub-scenario where the landing process is shaved off. Since it only focuses on the flying procedure of targeting in the mid-air, we could analyze this stage more precisely than the pointing task consisting of take-off and landing. In this section, we analyze the results of crossing tasks in order to better understand the properties of two types of controllers.

\subsubsection{Completion Time} 
We draw the linear regression between the difficulty index and completion time of all the crossing tasks, according to Fitt's Law. Figure \ref{fig:crossing-fitts-law-all} shows that the SOTA one-handed controller consumes less completion time in all the crossing tasks than the two-button controller. This result shares the same trend as ~\cite{arthur-drone}. Note that the slope of the two-button controller is larger than the one-handed controller, which indicates that the two-button controller is more sensitive to task difficulty. In contrast, even if the index of difficulty grows from 2.6 to more than 3.6, the one-handed controller's completion time slightly increases by less than one second. With VRFlightSIM, we could compare the completion time of different controllers, and observe the controllers' performance patterns on different tasks.

\subsubsection{Drone States} 
\begin{figure*}[h]
     \centering
    \begin{subfigure}[]{0.24\textwidth}
    \centering
    \includegraphics[width=\textwidth]{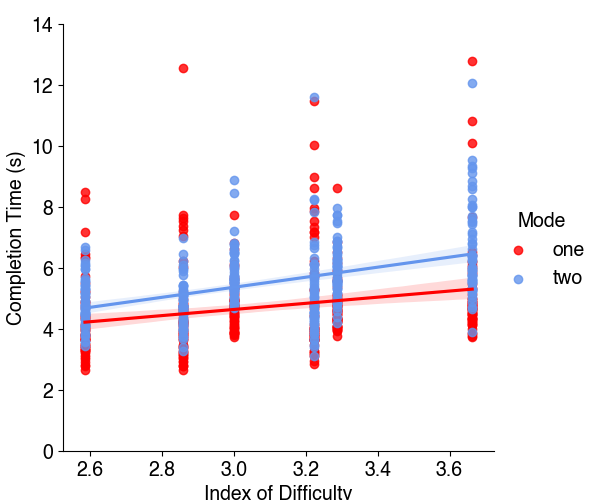}
    \caption{Completion Time of Crossing Tasks.}
    \label{fig:crossing-fitts-law-all}
\end{subfigure}
\hfill
     \begin{subfigure}[]{0.24\textwidth}
         \centering
         \includegraphics[width=\textwidth]{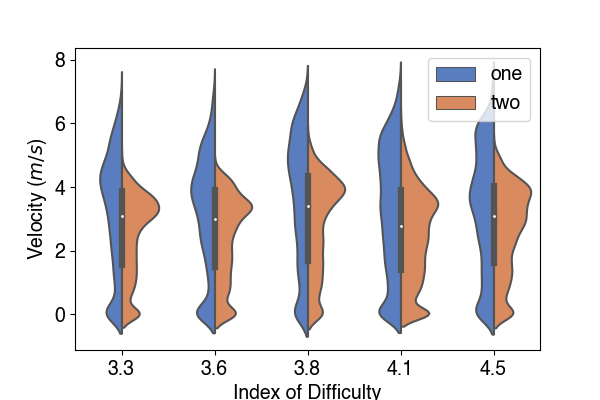}
         \caption{Violinplot of Velocity}
         \label{fig:crossing-velocity}
     \end{subfigure}
     \hfill
     \begin{subfigure}[]{0.24\textwidth}
         \centering
         \includegraphics[width=\textwidth]{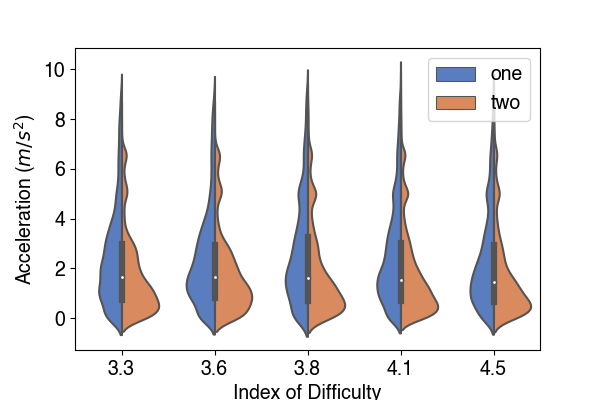}
         \caption{Violinplot of Acceleration}
         \label{fig:crossing-acceleration}
     \end{subfigure}
     \hfill
     \begin{subfigure}[]{0.24\textwidth}
         \centering
         \includegraphics[width=\textwidth]{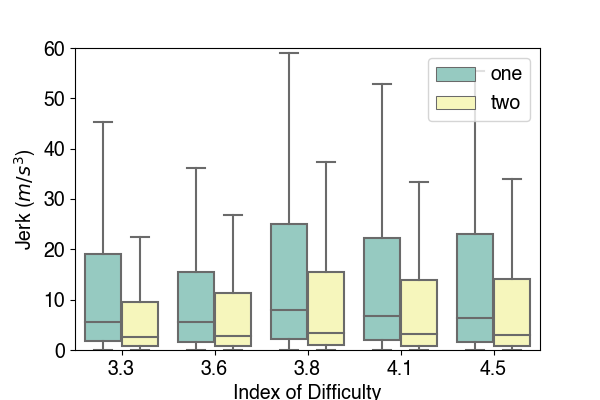}
         \caption{Boxplot of Jerk}
         \label{fig:crossing-jerk}
     \end{subfigure}
        \caption{Drone States of Crossing Task with Different IDs.}
        \label{fig:crossing-drone-states}
\end{figure*}
VRFlightSIM can capture the real-time drone states in the experimental procedures, including drone positions, velocity, acceleration, etc. By analyzing the drone states of each task, we can learn more details about the participant's performance and user behaviors 
with different controllers. Specifically, Figure \ref{fig:crossing-drone-states} depicts 
the drone states from three perspectives: velocity, acceleration, and jerk. First, in terms of velocity, Figure \ref{fig:crossing-velocity} shows 
the kernel estimation of the velocity distribution of two controllers on different tasks. The one-handed controller has a peak at higher flying speeds than the counterpart of the two-button controller, which means the participants tend to fly the drone faster with the one-handed controller. This corresponds to a lower task completion time. 

Second, in terms of acceleration, Figure \ref{fig:crossing-acceleration} offers the kernel estimation of real-time acceleration, where the one-handed controller has a higher peak value as well. Together with the velocity distribution, we can conclude that users are more confident at handling a one-handed controller for crossing operations, and meanwhile 
chasing a high speed to complete the task faster. 
Third, in terms of jerk values (the first time derivative of acceleration, which has been used to evaluate the comfort level of vehicles), Figure \ref{fig:crossing-jerk} is the boxplot of the real-time jerk of the flying drone. We find the potential risk 
in the one-handed controller, because of its 
higher jerk value than the two-button controller among all the tasks. A high jerk value means that the drone flew unstably with severe speed changes. Since people may use drones for photography purposes, 
the one-handed controller may add more disturbance, while the two-button controller might serve as 
a better choice because of the 
jerk value at a modest level. By analyzing the real-time drone states captured by VRFlightSIM, we can explore the characteristics of different controllers with in-depth analytics.

\subsubsection{Trajectory Patterns} 
\begin{figure}[h]
    \centering
    \includegraphics[width=\linewidth]{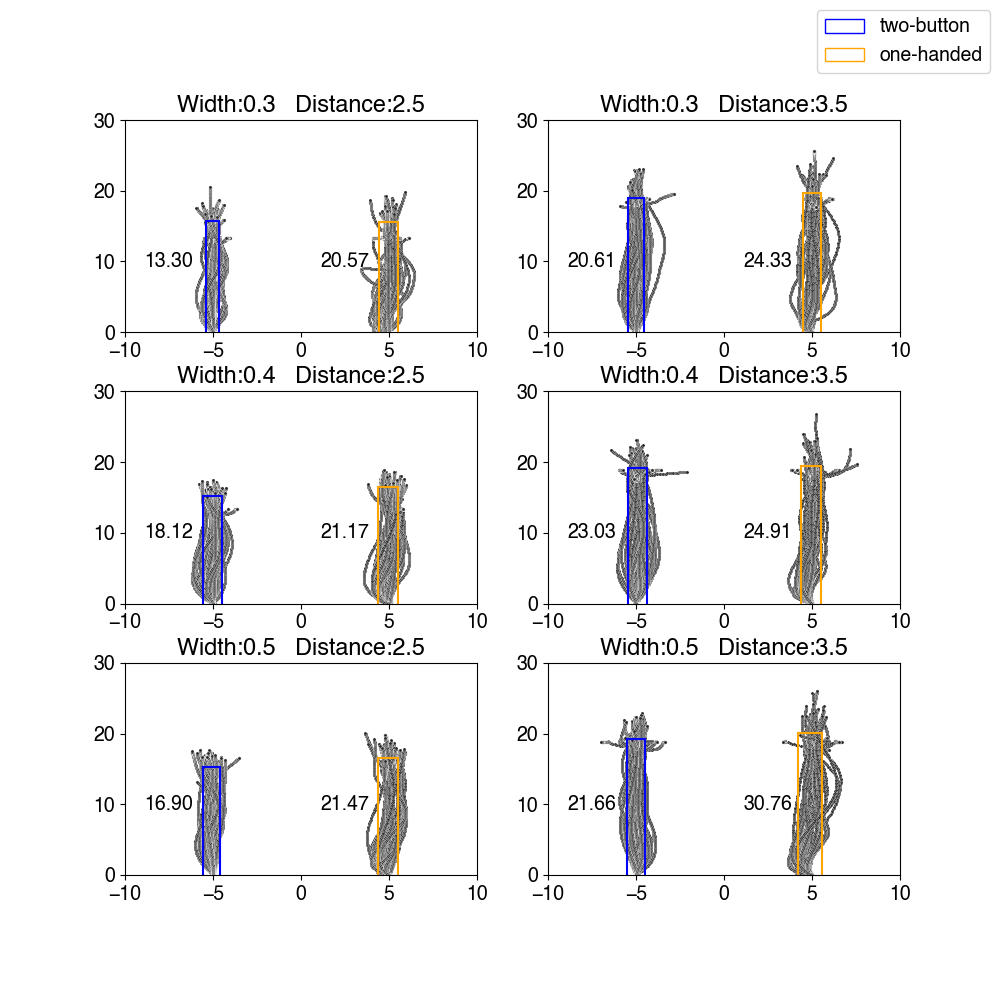}
    \caption{Drone's Trajectories on Crossing Tasks. The number near the rectangle denotes its area.}
    \label{fig:crossing-trajectory}
\end{figure}
Besides the task completion time and drone states, researchers might also be interested in drone's flying trajectory. Since VRFlightSIM can collect the real-time position of the drone, we can conveniently draw the flying trajectory. Figure \ref{fig:crossing-trajectory} illustrates all the trajectories of crossing tasks by different participants. We can compare the trajectory patterns among the two controllers. 
For almost all the tasks, the one-handed controller's trajectory is more dispersed with many longer twisted lines. To be more precise, we compute the standard deviation on each axis. The trajectory area corresponds to three times the standard deviation from the center (the blue/orange-colored squares).The smaller the trajectory area is, the more stably participants could complete the tasks with the corresponding controller. We note that the two-button controller has a smaller trajectory area than one-handed controller. The finding indicates that the participants could control the drone in a more normative scope without excessive manipulation, for example, overflying the drone away from the target. This insight is not captured by recent work~\cite{arthur-drone} because the previous work only deploy their experiments in a physical lab, where the drone positions are hardly being collected. 
With the convenient data capture by VRFlightSIM, we notice some weaknesses of the SOTA one-handed controller.

\begin{table*}[h]
\centering
\begin{tabular}{|c|cc|cc|cc|cc|cc|}
\hline
\textbf{Metric}    & \multicolumn{2}{c|}{\textbf{Completion Time}}                 & \multicolumn{2}{c|}{\textbf{Velocity}}                        & \multicolumn{2}{c|}{\textbf{Acceleration}}                    & \multicolumn{2}{c|}{\textbf{Jerk}}                            & \multicolumn{2}{c|}{\textbf{Trajectory Area}}                 \\ \hline
\textbf{Factor}    & \multicolumn{1}{c|}{\textbf{p-value}} & \textbf{F-statistics} & \multicolumn{1}{c|}{\textbf{p-value}} & \textbf{F-statistics} & \multicolumn{1}{c|}{\textbf{p-value}} & \textbf{F-statistics} & \multicolumn{1}{c|}{\textbf{p-value}} & \textbf{F-statistics} & \multicolumn{1}{c|}{\textbf{p-value}} & \textbf{F-statistics} \\ \hline
\textbf{Mode}      & \multicolumn{1}{c|}{7.81E-18}         & 77.7359               & \multicolumn{1}{c|}{4.8431E-14}       & 60.3394               & \multicolumn{1}{c|}{0.0015}           & 10.1788               & \multicolumn{1}{c|}{0.0446}           & 4.0557                & \multicolumn{1}{c|}{1.1953E-10}       & 43.0379               \\ \hline
\textbf{ID}        & \multicolumn{1}{c|}{5.55E-26}         & 27.8621               & \multicolumn{1}{c|}{1.5873E-47}       & 58.4736               & \multicolumn{1}{c|}{0.0004}           & 4.5486                & \multicolumn{1}{c|}{0.0046}           & 3.4353                & \multicolumn{1}{c|}{3.1615E-34}       & 38.6268               \\ \hline
\textbf{Mode * ID} & \multicolumn{1}{c|}{6.08E-05}         & 5.4555                & \multicolumn{1}{c|}{0.6774}           & 0.6294                & \multicolumn{1}{c|}{0.3330}           & 1.1500                & \multicolumn{1}{c|}{0.0112}           & 2.9981                & \multicolumn{1}{c|}{0.2842}           & 1.2502                \\ \hline
\end{tabular}
\caption{Two-way ANOVA analysis of Pointing Tasks: DoF = (1,5). All the Controller Interfaces (Mode) and Index of Difficulties (ID) passed the threshold of p-value$<$0.05.}
\label{table:pointing-anova}
\end{table*}

\subsection{Pointing}


\subsubsection{Completion Time}
\label{sec:completion-time}

\begin{figure*}[ht!]
     \centering
     \begin{subfigure}[b]{0.24\textwidth}
         \centering
         \includegraphics[width=\textwidth]{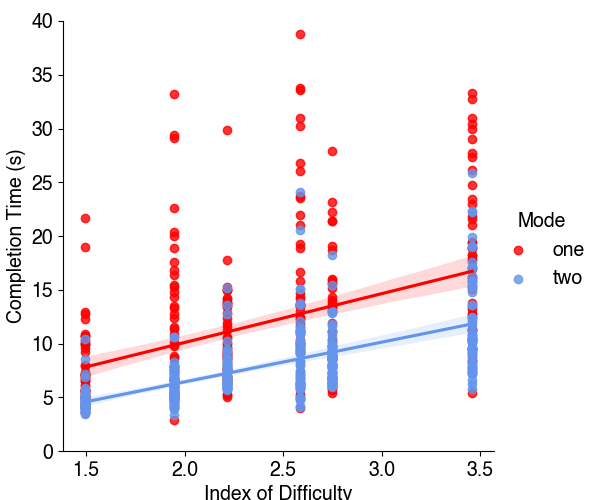}
         \caption{All}
         \label{fig:pointing-fitts-law-all}
     \end{subfigure}
     \hfill
     \begin{subfigure}[b]{0.24\textwidth}
         \centering
         \includegraphics[width=\textwidth]{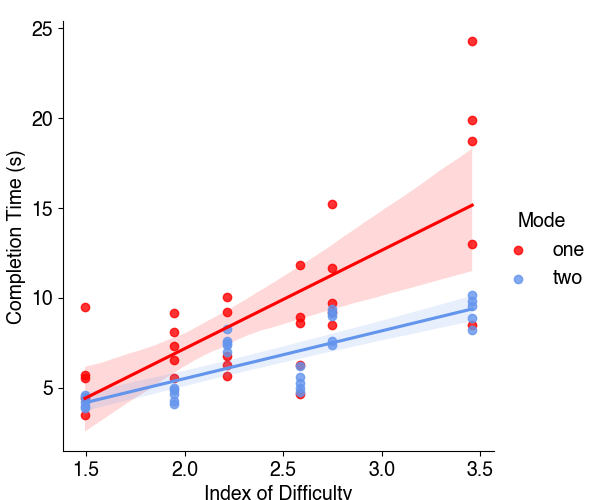}
         \caption{Two-button-preferred}
         \label{fig:fitts-law-type-1}
     \end{subfigure}
     \hfill
     \begin{subfigure}[b]{0.24\textwidth}
         \centering
         \includegraphics[width=\textwidth]{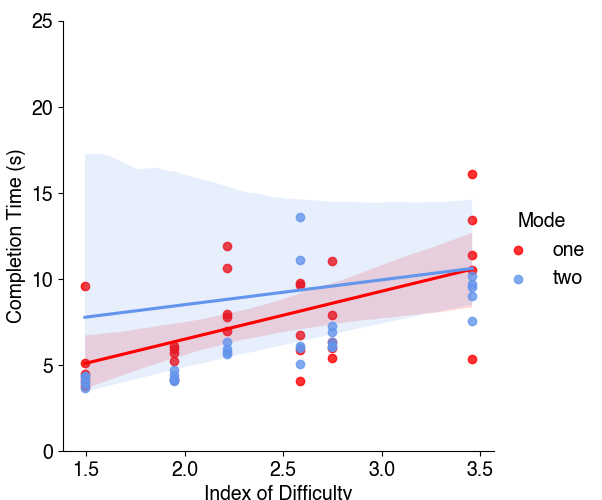}
         \caption{One-handed-preferred}
         \label{fig:fitts-law-type-2}
     \end{subfigure}
     \hfill
     \begin{subfigure}[b]{0.24\textwidth}
         \centering
         \includegraphics[width=\textwidth]{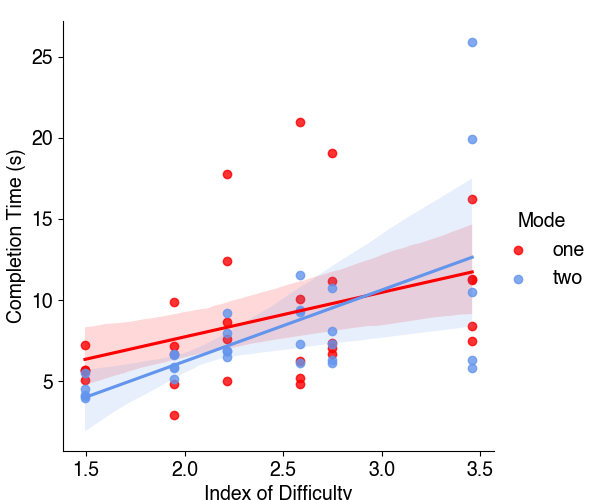}
         \caption{Hybrid (Neutral)}
         \label{fig:fitts-law-type-3}
     \end{subfigure}
        \caption{Three User Types for Pointing Task.}
        \label{fig:pointing-fitts-law-type}
\end{figure*}


As shown in Figure \ref{fig:pointing-fitts-law-all}, it can be clearly seen that the SOTA one-handed controller, in general, takes longer times of completing the same pointing task than the two-button controller.
Since the red line holds a higher slope (one-handed) than the blue line (two-button), it could be seen that the one-handed controller is more sensitive to tasks' difficulties. Note that this result is opposite to that of the crossing task. We attribute this phenomenon to the extra landing procedure of the pointing task, where the two-button controller surpasses the one-handed controller. With VRFlightSim, we can easily conclude that the traditional method of two-button controllers offers easier access for users on pointing tasks, especially for the drone landing subprocess.

However, if we examine the performance of individual participants, VRFlightSim gives a varied design implication of controller interfaces. 
We draw the same correlation graphs for participants with different preferences, in which 
their preferred controllers and hence behavior patterns could be mainly divided into three categories: (T1) Two-button-preferred (Figure \ref{fig:fitts-law-type-1}); (T2) one-handed-preferred (Figure \ref{fig:fitts-law-type-2});  (T3) Neutral (Figure \ref{fig:fitts-law-type-3}). 
Although provided with the same kinds of controllers dealt with the same tasks, user's learning curve is different from each other. For users of T1, their behavior pattern is the same as Figure \ref{fig:pointing-fitts-law-all}, where the two-button controller always performs better than the one-handed controller. For users of T2, the opposite observation 
happens -- 
the completion time of the two-button controller grows faster with the task difficulty, 
which demonstrates the sensitivity of these users to the task difficulty and the preference for the one-handed controller. 

Regarding users of T3, there is no apparent difference between the two types of controllers, so they acquire the skills of these controllers to the same extent. It is important to note that the intersection between the red and blue lines could imply design cues (More details are available in the next section. With VRFlightSim, we can easily notice the individual preference, as well as 
their learning pattern to decide the suitable controller design for scenarios of various difficulties. 

\subsubsection{Drone States}
\label{sec:drone-states}
\begin{figure*}[!ht]
     \centering
     \begin{subfigure}[b]{0.325\textwidth}
         \centering
         \includegraphics[width=\textwidth]{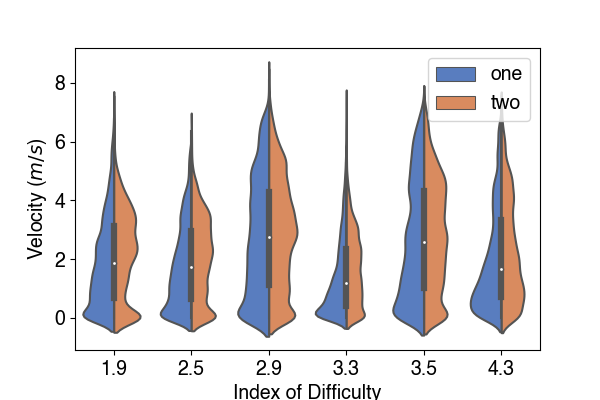}
         \caption{Violinplot of Velocity}
         \label{fig:pointing-velocity}
     \end{subfigure}
     \hfill
     \begin{subfigure}[b]{0.325\textwidth}
         \centering
         \includegraphics[width=\textwidth]{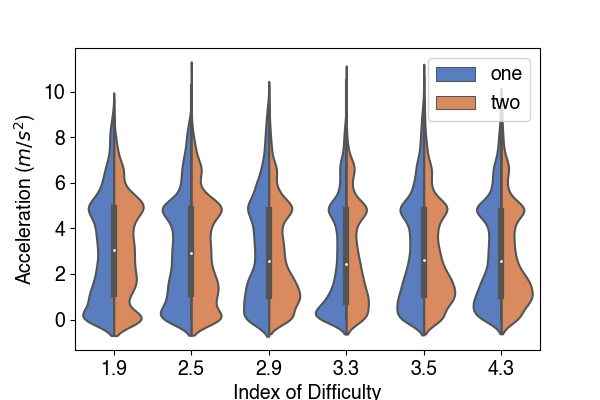}
         \caption{Violinplot of Acceleration}
         \label{fig:pointing-acceleration}
     \end{subfigure}
     \hfill
     \begin{subfigure}[b]{0.325\textwidth}
         \centering
         \includegraphics[width=\textwidth]{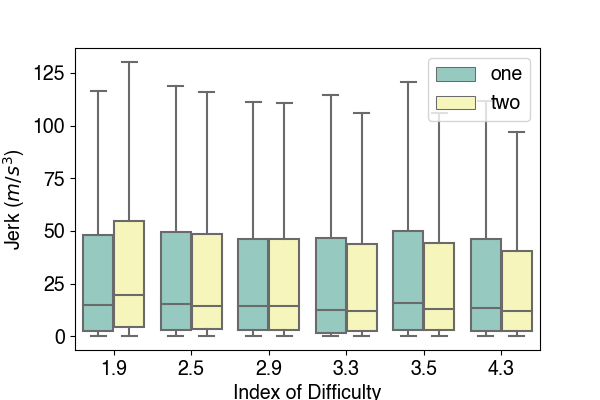}
         \caption{Boxplot of Jerk}
         \label{fig:pointing-jerk}
     \end{subfigure}
        \caption{Drone States of Pointing Task with Different IDs.}
        \label{fig:pointing-drone-states}
\end{figure*}
Same as the previous analysis on drone states for the crossing task, we try to study the drone's velocity, acceleration and jerk for the pointing task. Figure \ref{fig:pointing-velocity} describes 
the violin plot of the drone's velocity distribution under tasks with different IDs of task difficulties. Overall, participants tend to fly the drone faster using the two-button controller, revealing that 
they are more confident with the 
two-button controller for the sake of 
high speeds. 

About the drone's acceleration, in Figure \ref{fig:pointing-acceleration}, we can see that, the distribution of acceleration is poly-modal, where the peak at the zero value means a drone spends much time flying smoothly or staying still. We notice that the blue plots have lesser non-zero values among all the tasks, revealing that a drone receives less movement correction, commanded by users with the one-handed controller. In particular, 
we compute the real-time jerk value and Figure \ref{fig:pointing-jerk} states that, when handling easy tasks (i.e., when ID value is 1.9), a drone has less discontinuous acceleration with the one-handed controller, which means a more stable flying posture. However, as the task becomes difficult, the two-button controller is more likely to keep the drone's stability at a modest level. We could see that the one-handed controller has its advantage when handling easier tasks. This trend adds more details on top of 
Figure \ref{fig:pointing-fitts-law-all} that solely considers the completion time, but does not capture the drone's movement consistency that could impact various drone missions, e.g., photo and video taking.


\subsubsection{Trajectory Pattern}
\label{sec:trajectory-pattern}
\begin{figure}[h]
    \centering
    \includegraphics[width=\linewidth]{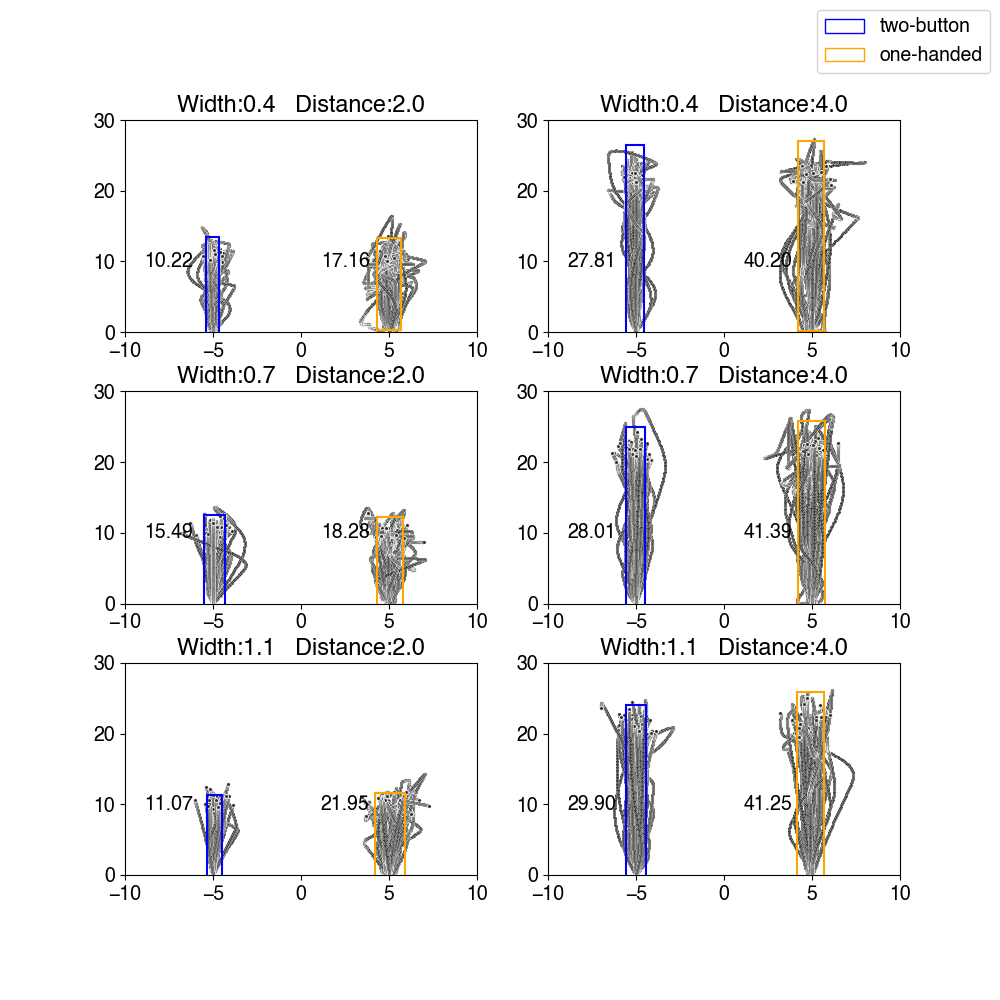}
    \caption{Drone's Trajectories on Pointing Tasks. The number near the rectangle denotes its area.}
    \label{fig:pointing-trajectory}
\end{figure}

Similar to the analysis on crossing tasks, we draw the trajectory of drone movement controlled by two kinds of controllers, to see whether the drone behaves stably in each experimental task. Figure \ref{fig:pointing-trajectory} illustrates the drone's moving trajectories under different pointing tasks, controlled by one-handed and two-button controllers. For all the six tasks, the two-button controller could help the participants achieve a significantly smaller drone trajectory area than the one-handed controller. This demonstrates that participants might act more stably when using the two-button controller, requiring fewer efforts in route correction and fewer unnecessary drone movements. This result is the same as that of crossing tasks.

\begin{figure}[!ht]
     \centering
     \begin{subfigure}[b]{0.48\linewidth}
         \centering
         \includegraphics[width=\linewidth]{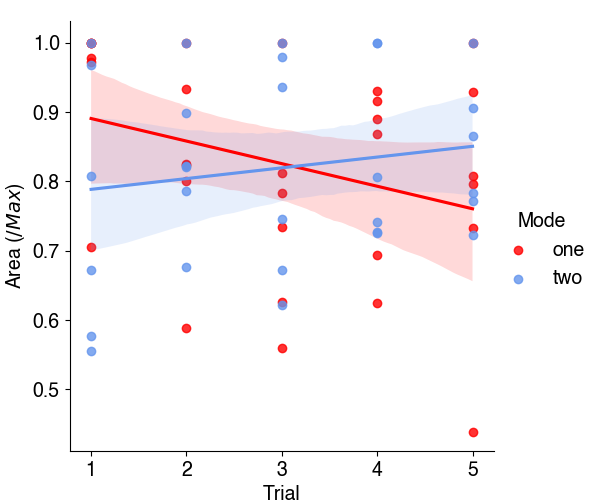}
         \caption{Crossing Tasks}
         \label{fig:crossing-trajectory-area}
     \end{subfigure}
     \hfill
     \begin{subfigure}[b]{0.48\linewidth}
         \centering
         \includegraphics[width=\linewidth]{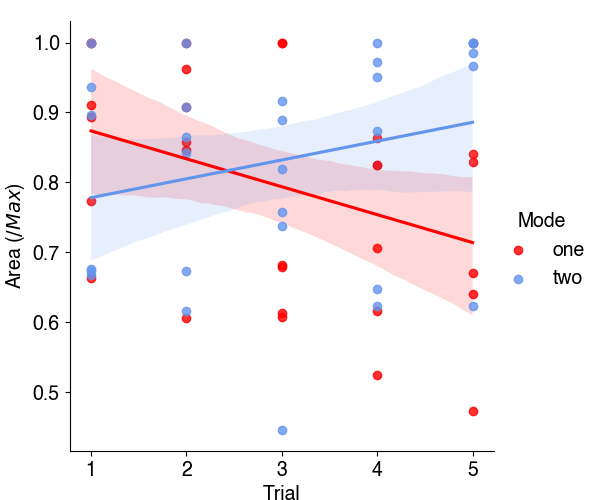}
         \caption{Pointing Tasks}
         \label{fig:pointing-trajectory-area}
     \end{subfigure}
        \caption{Trajectory Area by Trials.}
        \label{fig:pointing-drone-states}
\end{figure}

However, from a dynamic perspective, we notice some surprising results. Figures \ref{fig:crossing-trajectory-area} and \ref{fig:pointing-trajectory-area} illustrate the average trajectory area (take the relative value of the maximum) under different tasks by trials. 
As more trials are processed with the one-handed controller, participants are expected to be more skilled at handling the drone stably. Also, we anticipated that the trajectory area would become smaller. The trend was different when the participants ran the trials with the two-button controller. It can be seen that the trajectory area becomes more prominent with more and more trials. This phenomenon might reveal potential evidence of user behaviors. 
Once the participants become proficient and confident at handling the two-button controller, 
they tend to act more unrestrained and aggressive in the remaining trials, 
albeit they have been told to 
complete the tasks as stable as possible. With VRFlightSim, we could quickly notice such vulnerability in the early design stage, thus providing some advice for the users.

\section{Concluding Notes and Discussion}

In this paper, we designed and implemented an evaluation methodology with an open-source virtual environment (Microsoft Airsim), named VRFlightSim, to examine alternative input methods (i.e., new controllers) for drone flights. VRFlightSim aims to improve the ease of collecting drone performance data, such as completion time, flying trajectory, velocity, acceleration, and jerk, with flying diversified tasks, albeit in virtual environments. 

Identifying a large number of potential candidates of input methods with real-life experiments of drone flights are often tedious and expensive, while our methodology lowered the hurdle by offering (1) a preliminary understanding of flight controllers, and (2) virtual drones and environments as a comparable yet open testing ground that can potentially accumulate data from various testers/experiments. 
We note that VRFlightSim can screen out the inappropriate candidates, and work complementary with the real-life experiments, as a preliminary step in the entire cycle of interaction design. More importantly, the interaction designers have to conduct real-life experiments to test the effect of physical form factors caused by the actual devices, e.g., moving from a smartphone's touchscreen to a ring-form device~\cite{arthur-drone}. 

We further demonstrated the necessity of VRFlightSim with two flying tasks, namely pointing and crossing. We followed the proposed control method in~\cite{arthur-drone}, and re-implemented two control methods for a drone. VRFlightSim reveals a new profile of the one-handed control method driven force-assisted input. As reported in~\cite{arthur-drone}, their tasks show fixed difficulty ID, e.g., moving towards a direction (e.g., upward) for a 2-meter distance. In contrast, virtual environments can quantitatively capture the difficulty of flying tasks. Accordingly, VRFlightSim facilitates the construction of predictive models~\cite{d53-10.5555/2501707}. 

Furthermore, as indicated by the Fitts's Law, the task difficulty allows interaction designers to investigate the robustness of their proposed input methods in various scenarios, and further offers insights of mapping the controller designs with tasks. For instance, the intersection between two lines could indicate that a controller works better with certain tasks. 
As highlighted by the performance time during the pointing task, the user population with a higher preference to the one-handed controller (Figure~\ref{fig:fitts-law-type-2}) reflects that the one-handed controller can outweigh the two-button controller until it encounters tasks over 3.50 difficulty. Interaction designers with these cues can define that the highly mobile controller, driven by one-handed input with force levels, is suitable for easier tasks.
Furthermore, the users with a neutral preference for both methods (Figure~\ref{fig:fitts-law-type-3}) show the intersection at the task difficulty index value of 3.00. As such, interaction designers can narrow down the interaction scenarios, e.g., short-distance drone flights to landing targets of middle to large sizes. 

Human-drone interaction requires a common ground for evaluation. 
That is, the existing prototypes of drone controllers were experimented with under different conditions and external factors 
that led to difficult comparisons between studies. 
Re-building the testing environments or re-implementing the input methods are challenging. It is worthwhile to mention that a recent work~\cite{arthur-drone} limits to simple flying tasks and lacks data collection channels for drone movements (e.g., trajectory, jerk, acceleration). Otherwise, setting up numerous markers in room-scale environments to track the trajectories for such data collection is costly and time-consuming~\cite{pinpointfly}, not to mention the efforts of building physical obstacles~\cite{Yamada2019ModelingDC}. Thus, our paper calls for research efforts leveraging a common virtual platform of drone flight controls. 
Researchers can collaboratively employ our evaluation methodology\footnote{\url{https://github.com/alpha-drone-control/VRFlightSim}} to compare diversified controllers under a shared testing ground. 

\bibliographystyle{IEEE/IEEEtran}
\bibliography{sample-base}



\end{document}